\documentstyle[12pt]{article}
\def\be{\begin{equation}}
\def\ee{\end{equation}}
\def\ben{\begin{displaymath}}
\def\een{\end{displaymath}}
\def\ba{\begin{array}{c}}
\def\ea{\end{array}}
\begin{document}

\titlepage
\vspace*{2cm}

\begin{center}{\large \bf
Energy-dependent Hamiltonians and their
 pseudo-Hermitian interpretation

}\end{center}

\vspace{10mm}

\begin{center}
Miloslav Znojil,
  Hynek B\'{\i}la
    and V\'{\i}t Jakubsk\'{y}
\vspace{3mm}

\'{U}stav jadern\'e fyziky AV \v{C}R, 250 68 \v{R}e\v{z},
Czech Republic\\

e-mail: znojil@ujf.cas.cz

\end{center}

\vspace{5mm}

\section*{Abstract}

The concept of energy-dependent forces in quantum mechanics is
re-analysed.  We suggest a simplification of their study via the
representation of each self-adjoint and energy-dependent
Hamiltonian $H=H(E)$ with real spectrum by an auxiliary
non-Hermitian operator $K$ which remains energy-independent.
Practical merits of such an approach to the Schr\"{o}dinger
equations with energy-dependent potentials are illustrated using
their quasi-exactly solvable example.

\vspace{5mm}

PACS 03.65Ge

\newpage

\section{Introduction}

An introduction of phenomenological potentials which vary with
the energy of the system proved useful in atomic, molecular as
well as nuclear and particle physics. Quite often, a certain
energy-dependence of the parameters results from their
physical meaning -- the best known illustrations are provided by
the Klein-Gordon equation (where the potential happens to be a
quadratic function of the energy \cite{Landau}) and by the
Bethe-Salpeter equations (where the relativistic kinematics is
partially incorporated in the description of the two-body
systems \cite{BS}). Another elementary sample of such a
motivation of the work with an energy-dependent Hamiltonian
$H=H(E)$ may be found outlined here in Appendix A.

In another group of phenomenological models, the
energy-dependence in the ``realistic" Hamiltonians $H=H(E)$
remains artificial and serves as a mere {\it ad hoc} simulation
of certain complicated and not too well understood effects
within a sufficiently transparent model.  In the similar
pragmatic simulations, the fit of the experimental data is often
unexpectedly successful~\cite{Hirota}.  Although the consequent
theoretical foundation of the {\em explicit construction} of the
energy-dependent components of the interactions may be missing,
the procedure of the (variational or numerical) fitting itself
usually offers a useful guidance towards an optimal choice of
the form of $H=H(E)$~\cite{HirotaII}.  In such a situation, one
can only regret that the merits of the simplified physics are
sometimes counterbalanced by perceivable growth of mathematical
difficulties. In this sense, the energy-dependent forces attract
attention as a purely mathematical challenge~\cite{genII}.

From an abstract point of view, one of the most common
theoretical sources of the energy-dependent potentials may be
sought in a tentative projection of wavefunctions $|\Psi\rangle$
on a certain ``model" subspace of Hilbert space. After such a
projection the unitarity of the time evolution is broken of
course. The microscopic and energy-independent Hamiltonians $H$
are being replaced by their so called effective reduced forms
$H_{eff}=H_{eff}(E)$ \cite{Feshbach}. Some of the key technical
details of such a re-definition of the system may be found
summarized in Appendix B.  Within its overall framework, the
development of a sufficiently flexible treatment of specific
models seems to be an urgent task. For this reason, we decided
to re-analyze here the general Schr\"{o}dinger energy-dependent
bound-state problem
 \be
H(E_\alpha)|\phi_\alpha\rangle
=E_\alpha\,|\phi_\alpha\rangle\,
\label{SE}
 \ee
keeping in mind, first of all, the new possibilities which were
opened by the recent growth of interest in the systems where the
time evolution need not necessarily be purely
unitary~\cite{pseudo}.

In section 2 we shall start our considerations by summarizing a
few known facts about the models with the general $H = H(E)$. We
shall show there how the major part of the difficulties
introduced by the energy-dependence of the Hamiltonian may be
formally eliminated by the introduction of a suitable
bi-orthogonal basis. In the new language, the enhanced
flexibility of the basis becomes reflected by the possibility of
a modification of the scalar product in Hilbert space. In this
manner, one may even try to control the physical contents of the
theory~\cite{annals}.

In  section 3 we shall pay attention to the demonstration of the
applicability of the formalism of section 2 to a specific class
of the energy dependent Hamiltonians.  We decided to pay
attention to the next-to-solvable potentials which form the so
called quasi-exactly solvable (QES, \cite{Ushveridze}) family. A
sketchy review of their basic properties is presented as a
separate Appendix C.  The main consequence of our present
energy-dependent re-interpretation of the popular QES
Hamiltonians may be seen in the establishment of their new role
of a source of non-orthogonal bases in Hilbert space.

In the short summary of our effort in section 4 we shall emphasize
that our Appendices A and B stressed the immediate physical and
pragmatic appeal of the energy-dependence in $H=H(E)$. Our subsequent
choice of the QES illustrations of Appendix C may then play the role
of their sophisticated and more flexible mathematical complement,
showing that although our present interpretation of the effective and
other energy-dependent Hamiltonians seems to be a fairly universal
recipe, its various implementations may differ in the degree of the
insight they are able to offer.  In this sense, any continuation of
their study would be welcome.

\section{Mathematical consequences of the energy-dependence of
$H(E)$}

In the spirit of the recent review \cite{Mares}, many (though
not all of the) mathematical puzzles related to eq. (\ref{SE})
may be clarified when one contemplates an auxiliary family of
the Hermitian Hamiltonians $H(z)$ which depend on a real
parameter $z$. Let this value be temporarily disentangled from
the energy eigenvalues. Then, all the Hamiltonians $H=H(z)$ may
be assigned their respective spectral representations,
 \ben
H(z) =\sum\,|n(z)\rangle
\,E_n(z)\,\langle n(z)|\,.
\een
Only within such a broader framework, we impose the constraint
$E_n(z) = z$ at the very end of all the necessary constructions
and for all the indices $n$.  The latter constraint may happen
to be satisfied at an empty or non-empty set of roots $z=
z_1(n), z_2(n),\ldots, z_{m(n)}(n)$. With their knowledge at our
disposal (see an elementary example in Appendix A), one may pick
up a suitable subset $A$ of the ``allowed" multiindices
$\alpha=(n,i) \in A$ and abbreviate $z_i(n)=E_\alpha$ and
$|n(z_i\rangle = |\phi_\alpha \rangle$. By such a construction
we arrive at a formal solution of our energy-dependent problem
(\ref{SE}) under very general assumptions and/or for the various
particular choices of the parameter-dependence in $H=H(z)$.

\subsection{Non-orthogonal wave functions}

There exist several difficulties which arise in connection with
the generalized Schr\"{o}dinger equation (\ref{SE}).  First of
all, the standard orthogonality relations between the separate
bound states are lost. Even though the norm of each state may be
fixed via a suitable re-scaling of $ ||\phi_\alpha||= \langle
\phi_\alpha |\phi_\alpha \rangle $, it is necessary to know and
evaluate also all the off-diagonal non-vanishing overlaps
 \be
 \langle \phi_\alpha |\phi_ \beta \rangle =R_{\alpha, \beta},
\ \ \ \ \ \ \alpha, \beta \in A\,. \label{overlaps}
 \ee
Under certain (nontrivial) assumptions this matrix has an
inverse which enters the decomposition of the identity projector
 \be
\hat{I} = \sum_{\alpha, \beta\in A}\, |\phi_\alpha \rangle \left (
R^{-1} \right )_{\alpha, \beta}
 \langle \phi_ \beta|\,.
 \label{CR}
 \ee
The practical value of such a completeness of our basis (or, at
least, its completeness in the whole ``relevant" Hilbert space) is
significantly lowered by the non-diagonality of the matrix $R$. The
sufficiently precise evaluation of this matrix {\em and} of its
inverse $R^{-1}$ may happen to become {\em prohibitively}
time-consuming.  The practical applicability of the completeness
relations (\ref{CR}) with a non-diagonal $R$ is further lowered in
all the applications of the energy-dependent potentials which would
rely upon a purely numerical solution of eq. (\ref{SE}).  Within the
framework of our present considerations, at least a partial return to
semi- and non-numerical techniques will be advocated, therefore.

\subsection{Biorthogonal basis in the Hilbert space}

An inspection of the action of our {\em family of the operators}
$H(E)$ on the separate elements of our basis set of kets
$|\phi_\alpha \rangle \equiv |n(z_i\rangle $ with  $\alpha=(n,i)
\in A$ reveals that the new operator $K$ defined by its
generalized spectral representation
 \be
K
=
\sum_{\alpha, \beta \in A}\, |\phi_\alpha \rangle \,E_\alpha\, \left
( R^{-1} \right )_{\alpha, \beta} \langle \phi_ \beta| \,
 \label{Ka}
 \ee
may be interpreted as equivalent to our Hamiltonian family
whenever their action within our ``allowed" space is concerned,
 \be
K\,|\phi_\alpha\rangle =E_\alpha\,|\phi_\alpha\rangle\,. \label{SEK}
 \ee
By construction, the new operator $K$ is energy-independent
(i.e., no nonlinearity is encountered) and non-Hermitian (this
is the price we decided to pay). The latter property of $K \neq
K^\dagger$ is unexpected  but its origin is clear after we
abbreviate
 \be
\langle\langle \phi_\alpha| = \sum_{ \beta\in A}\, \left ( R^{-1}
\right )_{\alpha, \beta}
 \langle \phi_ \beta|\,
 \label{ufe}
 \ee
and re-interpret the above eq. (\ref{Ka}) as a diagonal-type
spectral representation of our auxiliary non-Hermitian
quasi-Hamiltonian,
 \be
K
=
\sum_{\alpha \in A}\,
|\phi_\alpha \rangle
\,E_\alpha\,\langle
 \langle \phi_\alpha|
\,.
 \ee
The simplicity and transparency of such a notation is based on the
consequent use of the double-bra symbol for the left eigen-vectors.
This convention introduces a basis which happens to be, by
construction, bi-orthogonal,
 \be
\langle \langle \phi_\alpha |\phi_ \beta \rangle =\delta_{\alpha,
 \beta}, \ \ \ \ \ \ \alpha, \beta \in A\,.
 \ee
Similarly, the abbreviated eq.~(\ref{CR}),
 \be
\hat{I} = \sum_{\alpha\in A}\, |\phi_\alpha \rangle \langle
 \langle \phi_ \beta|\,
 \label{CRA}
 \ee
represents the maximally compact form of completeness
relations.

\subsection{Pseudo-Hermiticity }

In a way paralleling the recent interest in certain specific
non-Hermitian (for example, ${\cal PT}-$symmetric) systems with
real spectra \cite{sample}, we may expect that there exists an
invertible and Hermitian ``auxiliary metric" operator
$\eta=\eta^\dagger$ such that our new operators $K \neq
K^\dagger$ will satisfy the following quasi- or
pseudo-Hermiticity relations
 \be
 K^\dagger \eta = \eta\,K\,
 \label{ref}
 \ee
\cite{pseudo}. After the insertion of an ansatz
 \be
\eta = \sum_{\alpha, \beta\in A}\, |\phi_\alpha \rangle \rangle\,
U_{\alpha, \beta} \,\langle
 \langle \phi_ \beta|\,,
 \label{CRe}
 \ee
relation (\ref{ref}) degenerates to the mere algebraic equation
 \be
E_\alpha U_{\alpha, \beta}=  U_{\alpha, \beta}\,E_ \beta, \ \ \ \ \ \
\ \alpha, \beta \in A\,.
 \ee
It implies that under the non-degeneracy assumption $ E_ \beta \neq
E_\alpha$ for $\alpha \neq  \beta$, all the off-diagonal elements of
$U$ must vanish while the diagonal elements $
U_{\alpha,\alpha}=d_\alpha=d_\alpha^*\neq 0$ remain variable. This
defines the large set of the metrics
 \be
\eta = \sum_{\alpha \in A}\, |\phi_\alpha \rangle \rangle\,
d_{\alpha} \,\langle
 \langle \phi_\alpha|\,
 \label{CRf}
 \ee
which are self-adjoint and invertible,
 \be
\eta^{-1} = \sum_{\alpha \in A}\, |\phi_\alpha \rangle \,
d_{\alpha}^{-1} \,
 \langle \phi_\alpha|\,.
 \label{CRg}
 \ee
In this language each energy-dependent Hamiltonian $H(E)$ may be
better understood via its $\eta-$pseudo-Hermitian representation
$K$ using an arbitrary auxiliary metric $\eta$ defined by eq.
(\ref{CRf}). Its role becomes clear after we introduce the
Hermitian conjugate of eq.  (\ref{SEK}), $\langle\phi_\alpha
|\,K^\dagger =\langle\phi_\alpha |\,E_\alpha^*$. At the real
energies $E_\alpha^*=E_\alpha$ it enables us to re-write this
equation in the spirit of eq.  (\ref{ref}),
 \ben
 \left (
 \langle \phi_\alpha|\,\eta
 \right )\,K =
 \left (\langle
 \phi_\alpha|\,\eta
 \right )\,E_\alpha\,.
 \een
The non-degeneracy assumption immediately implies that
 \be
 \langle \langle \phi_\alpha| = const\,
 \langle
 \phi_\alpha|\,\eta\,.
 \label{renomr}
 \ee
We see that the knowledge of the (non-singular) operator $\eta$
is, up to a re-normalization (\ref{renomr}), equivalent to the
knowledge of the matrix $R$ [cf. eq. (\ref{ufe})]. The main
meaning of the use of $\eta$ follows from the fact that in the
hypothetical Hilbert space where the metric would be given by
the operator $\eta$, the operator $K$ would become, in the light
of eq. (\ref{ref}), ``Hermitian".

\section{Illustration: Partially solvable potentials}

The Hautot's QES model (\ref{charged}) of Appendix C is an example of
the well-motivated choice of the quantum system where, by definition,
a suitable selection of the energy-dependent Hamiltonian $ H(E) =
H_{(F)}= H_{(0)} + F\,W$ [with $W = 1/r$ and $F=F^{(QES)}(E)$] is
{\em dictated} by the feasibility of the evaluation of the overlaps
(\ref{overlaps}). This means that the {\em selected} bound states
$\psi_{n,\ell}^{(QES)}(r)$ remain represented by polynomials of a
{\em finite} degree $N \geq n$. While we may drop the information
about the value of the nodal count $n$ and of the angular momentum
$\ell$ as redundant, we still have to keep in mind our choice of the
subscript $j$ which numbers all the eligible QES charges $e =
F_{N,j}$ with $j = 0, 1, \ldots, N$.  This allows us to compactify
our notation, $\psi_{n,\ell}\equiv \psi_{n,\ell,N,j} \to
|N,j\rangle$, with the main purpose of using a suitable subset of the
overcomplete family of these QES bound states as a non-orthogonal
basis in the physical Hilbert space.

\subsection{QES basis and orthogonality-type relations}

Let us re-write eq. (\ref{charged}) as the following set of
equations in the Dirac's bra-ket notation,
 \begin{equation}
  H_{(0)}
 \ |M,k\rangle
  +W\ |M,k\rangle\, F_{M,k}
 =  |M,k\rangle\,E_M
 \label{one}
 \end{equation}
 \[
 \ \ \ \ \ \ \ \ \ \ \  \ \ \ \
 \ \ \ \ \ \ \ \ \ \ \  \ \ \ \
 \ \ \ \ \ \ \ \ \ \ \  \ \ \ \
 k =  1,2, \ldots, {\it L}(M), \ \ \ M = 0, 1, \ldots \,.
 \]
In our particular example we shall prefer the
one-charge-per-energy choice of ${\it L}(M)=1$.  The left and
right eigenstates do not differ (or at least need not differ --
see the non-Hermitian generalization of eq.  (\ref{charged}) in
ref. \cite{pthau}).  Thus, we may complement eq. (\ref{one}) by
its conjugate counterpart
 \begin{equation}
 \langle N,j | \
 H_{(0)} + F_{N,j}\,  \langle N,j | W
 = E_N\ \langle  N,j |,
 \label{two}
 \end{equation}
 \[
 \ \ \ \ \ \ \ \ \ \ \  \ \ \ \
 \ \ \ \ \ \ \ \ \ \ \  \ \ \ \
 \ \ \ \ \ \ \ \ \ \ \  \ \ \ \
 j =  1, 2,\ldots, {\it L}(N), \ \ \ N = 0, 1, \ldots \,.
 \]
There is no {\it a priori} reason for an orthogonality between
multi-indexed bra vectors $ \langle N,k|\ \equiv\  \langle A|\ $ and
ket vectors $\ |N',k'\rangle\ \equiv\   |b\rangle \ $. Their overlaps
$R_{A,b}=  \langle A | b \rangle$ form a non-diagonal matrix
(\ref{overlaps}) in general. We only have to assume that this matrix
remains invertible. Only in such a case we may employ eq. (\ref{CR})
and introduce the identity projector
 \be
 I = \sum_{
 a \in {\it J}_{ket},
 B \in {\it J}_{bra}
 }
 |a\rangle\,
\left ( R^{-1} \right ) _{a,B}\, \langle B|\ .
 \label{projek}
 \end{equation}
As long as equations (\ref{one}) and (\ref{two}) share all their
eigen-energies and eigen-charges, we may write down the
following two alternative projected equations
 \begin{equation}
 {} \langle N,j | \
 H_{(0)}
 \ |M,k\rangle = {} \langle N,j |M,k\rangle\  E_{M}
 -
 {} \langle N,j | \
  W
 \ |M,k\rangle\ F_{M,k}
 \label{oned}
 \end{equation}
 \begin{equation}
 {} \langle N,j | \
  H_{(0)}
 \ |M,k\rangle = E_{N}\  {} \langle N,j |M,k\rangle
 - F_{N,j}\,
 {} \langle N,j | \
  W
 \ |M,k\rangle
 \label{twod}
 \end{equation}
with $ (N,j) \in {\it J}_{bra}$ and $ (M,k) \in {\it J}_{ket}$. Their
subtraction gives the constraint
 \begin{equation}
 \left (
 F_{M,k}-F_{N,j}
 \right )\,
 {} \langle N,j | \
  W\ |M,k\rangle
  =
 \left (
 E_{M}-E_{N}
 \right )\,
  R_{({N,j}),({M,k})}\,
  .
 \label{biogg}
 \end{equation}
This relation may be understood as the energy-dependence-related
generalization of the usual orthogonality of the eigenvectors.  All
the matrix elements of $H_0$ dropped out.  For the so called Sturmian
multiplets (i.e., within each subspace where $M = N$), the
left-hand-side expression must be a diagonal matrix with respect to
its second indices $j$ and $k$.  For our present purposes we shall
abbreviate ${}\langle N,j |  W |N,j\rangle \equiv w_{N,j}$.  All the
other matrix elements of $W$ may be treated as defined by eq.
(\ref{biogg}).  Once we know all the overlap matrix $R$, just the
diagonal elements $ w_{N,j}$ remain unspecified and, whenever needed,
their values must be generated by an independent calculation.

\subsection{Perturbations and non-QES states}

Whenever we leave the safe QES domain and contemplate the more
general bound-state problem (\ref{charged}) at a generic charge
$F\neq F^{(QES)}$, we encounter a non-elementary (i.e., {
numerical or perturbative}) diagonalization of Schr\"{o}dinger
equation
\[ \left [
H_{(0)} + F\,W \right ]\ |\Psi \rangle = E(F) \ |\Psi \rangle\,.
\label{obone},
\]
Assuming that $ F\neq F^{(QES)}$ and using eq.
(\ref{projek}) we may insert
\[ |\Psi\rangle= \sum_{ a \in {\it
J}_{ket}, B \in {\it J}_{bra} } |a\rangle\,\left ( R^{-1} \right
)_{a,B}\,{} \langle B|\Psi \rangle = \sum_{ a \in {\it J}_{ket}
} |a\rangle\,h_{a}\,
\]
and arrive at the infinite-dimensional linear algebraic equation
 \[
 \sum_{b \in {\it J}_{ket},
 C \in {\it J}_{bra}
 }
 \,{} \langle A | \,H(F)\,|b\rangle\,
 R_{b,C}\,{}\langle C|\Psi\rangle = E\,
 {}\langle A|\Psi\rangle, \ \ \ \ \ \ \ \ \ A \in {\it
 J}_{bra}
 \]
i.e., matrix equation
 \begin{equation}
 {\bf Z}(E,F) \,\vec{h} = 0, \ \ \ \ \
  \label{dvoje}
 \end{equation}
where
 \[
 {\bf Z}_{A,b}(E,F)
 = {} \langle A |\ H_{(0)}\ | b \rangle
 -E\, {} \langle A | b \rangle
 +F\, {} \langle A |\ W\ | b \rangle\ .
 \]
In the preparatory step we must evaluate {\em all} the input
matrix elements. This step is usually the most time-consuming
part of the algorithm.  Fortunately, it can quite efficiently be
optimized, in the present QES setting, by the use of all the
available orthogonality-type identities.  Thus, we recall eq.
(\ref{twod}) and eliminate all the matrix elements of $H_{(0)}$.
This means that in eq.  (\ref{dvoje}) the costly input
information becomes reduced to the mere evaluation of the matrix
elements of the Coulombic $W(r)=1/r$,
\[
 {\bf Z}_{A,b}(E,F)
 =
  \left ( F-F_A \right ) \, {} \langle A |\ W\ | b \rangle
 -\left ( E-E_A \right )\, {} \langle A | b \rangle
  \ .
 \]
In the second step we keep $M \neq N$ (i.e., we stay out of the
Sturmian subspaces or diagonal blocks in the matrix ${\bf Z}$) and
postulate the absence of a random degeneracy of charges. This means
$F_{M,k} \neq F_{N,j}$ so that we are permitted to re-arrange the
orthogonality-like relation (\ref{biogg}) into
definition representing a further vital reduction of the
necessity of the excessive
numerical integrations,
 \[
 {} \langle N,j | \
  W\ |M,k\rangle
  =
 \frac{
 E_{M}-E_{N}
 }{
 F_{M,k}-F_{N,j}
 }
  R_{({N,j}),({M,k})}\,, \ \ \ \ \ \ \ \ \ \ M \neq N\ .
 \label{bioggpp}
 \]
The final, maximally reduced form of our linear Schr\"{o}dinger
non-QES algebraic problem then reads
 \[
 w_{N,j}\,h_{N,j}+
 \sum_{K(\neq N),p}\,
  \frac{E_N-E_K}{F_{N,j}-F_{K,p}}\
 R_{(N,j),(K,p)}\,h_{K,p}   =
 \]
 \[
  \ \ \ \ \ \ \ =
 \frac{
 E-E_N
 }{
 F-F_{N,j}
 }\,
 \  \sum_{M,k}\,
 R_{(N,j),(M,k)}\,h_{M,k}\ ,
 \]
 \[
 \ \ \ \ \ \ \ \  \  \ \ \ \ \ \
 \ \ \ \ \ \ \ \
 \ \ \ \ \ \ \ \  \
 \ \ \ \ \ \ \ \  \
 j = 1, 2, \ldots, {\it L}(N), \ \ \ \
 N = 0, 1, \ldots\ .
 \]
Summarizing, any numerical or perturbative solution of this algebraic
system requires just the knowledge of the overlaps $R$ and of the
single array of the special diagonal Coulombic matrix elements
$w_{N,j}$.

\section{Summary}

Whenever we try to parallel the orthogonality proof as it works in
the standard energy-independent cases \cite{Landau}, we merely obtain
a very formal relation \cite{Mares}
 \be
\langle \phi_ \beta | \left [ H(E_ \beta)-H(E_\alpha)
 \right ]
 |\phi_\alpha\rangle
=
\left ( E_ \beta -E_\alpha \right )
 \,
\langle \phi_ \beta |\phi_\alpha\rangle\,.
 \label{SEW}
 \ee
In spite of its comparative weakness (and in spite of its virtually
negligible role in the purely theoretical considerations of section 2
above), we succeeded in demonstrating that its contents remain
non-empty and that its practical implications may be very useful.
First of all, this ``weak orthogonality" relation has been shown to
imply that whenever the energies coincide, $ E_ \beta =E_\alpha$, the
self-overlaps $\langle \phi_ \beta |\phi_\alpha \rangle$ remain
undetermined.  Less trivial is the observation that for the
non-degenerate spectra with $ E_ \beta \neq E_\alpha$, formula
(\ref{SEW}) admits the non-vanishing of the overlaps $R_{
\beta,\alpha}$ {\em and} relates their values to the left-hand-side
matrix elements.  This is one of the key consequences of eq.
(\ref{SEW}), with practical merits which vary with the explicit form
of $E-$dependence of the Hamiltonian operator $H(E)$. This point of
view has been more explicitly supported by our illustrative
energy-dependent re-interpretation of the current and popular QES
examples.

In the broader context involving the generic energy-dependence of
virtually all the effective Hamiltonians we re-interpreted all the
energy-dependent (i.e., in a way, non-linear) Schr\"{o}dinger
equations (\ref{SE}) as  equivalent to the non-Hermitian but, by
construction, fully linear algebraic eigenvalue problems (\ref{SEK}).
Such a step has been shown to faciliate our understanding of the
theory as well as of the deeper mathematical nature of our original
equation.  The related possible scalar-product re-interpretations may
make it more relevant in the context of physics, along the lines
discussed much more thoroughly in the review of quasi-Hermiticity
\cite{annals} as well as in its more recent continuations
\cite{pseudo,critique}.

In the future extensions of our present note one could select several
interesting directions.  Firstly, whenever the deviation $\lambda= F-
F^{(QES)}$ of the charges remains sufficiently small, one feels
tempted to construct the spectra $E= E(\lambda)$ perturbatively, in
the form of a power series in $\lambda$. This could complement the
existing QES-related perturbative studies based on different
principles \cite{perq}. In a broader setting, our present
recommendation of the use of unusual biorthogonal bases might also
lead to a direct new progress in the area of perturbation theory
itself.

As we already emphasized, the diagonalization of an operator which
depends on its own eigenvalues is not, strictly speaking, a linear
problem.  In this sense the incompletely elementary QES class seems
particularly suitable as an illustrative example at an introductory
stage. In a way emphasized in ref. \cite{Mares}, all continuations of
such a direction of development will be well motivated since even
many apparently elementary energy-dependent interactions fail to
admit a non-numerical treatment.  Thus, a systematic classification
of all the solvable cases would be of a paramount theoretical as well
as practical importance.

Last but not least, we should not forget that also the recently
popular replacements of the Hermitian $H=H(E)$ by their
non-Hermitian descendants with real spectra \cite{pthau} could
offer another inspiration for a continuation of our present
study.

\section*{Acknowledgements}

We all appreciate numerous inspiring discussions with Ji\v{r}\'{\i}
Form\'{a}nek, Roland Lombard and Ji\v{r}\'{\i} Mare\v{s}. Work
supported by GA AS \v{C}R, grant Nr. 104 8302.

 \newpage

\section*{Appendix A: Energy dependence
with an origin in physics}

Under the influence of experimental data, even such a certainty
as the energy-independence of the mass of a particle may require
a critical re-evaluation.  Such a critique finds its support in the
decays of mesons ${K}^+$ where, traditionally, the vector-meson
dominance offers a parameter-free explanation of the overall
character of the data \cite{Licharda}.  Still, the standard
choice of the form factor $F(t)$ of the $\rho$ meson leads to a
perceivable discrepancy between the parallel descriptions of the
decays ${K}^+ \to \pi^+\mu^+\mu^-$ and ${K}^+ \to \pi^+e^+e^-$.
The remedy of this discrepancy has been found \cite{Lichardb} in
the variability of the dilepton mass $M= t_{phys}$ which is
perceivably different in the above two processes.  In accord
with Isgur et al \cite{Isgur}, one must work with the
running mass $m_{\rho}^2\to \tilde{m}_{\rho}^2(t)$ in
 \be
 F(t) = \frac{m_{\rho}^2}{m_{\rho}^2-t}\ \ .
 \label{jedna}
 \ee
The variability of the running mass with the cross-channel
energy $t$ is necessary for a consistent interpretation of the
analyticity of the propagators.  Explicit calculations lead
to the further modification of eq. (\ref{jedna}) and an
imaginary shift appears in
 \be
 F(t) = \frac{\tilde{m}_{\rho}^2(0)}{\tilde{m}_{\rho}^2(t)
 -t-i\,{m}_{\rho}\,\Gamma(t)}\ \ .
 \label{jednabe}
 \ee
The determination of the ``realistic" running-mass function
$\tilde{m}_{\rho}^2(t)$ represents the main and challenging
theoretical problem \cite{Lichardc}. In the phenomenological
considerations of ref. \cite{Lichardb}, a satisfactory agreement
between the experiment and its description has been achieved via
a selfconsistent determination of the energy dependence of the
mass $\tilde{m}_{\rho}^2(t)$ near the phenomenologically
relevant dilepton invariant energies $t \approx E_{\mu^+\mu^-}$
and $t \approx E_{e^+e^-}$.

For a schematic clarification of some of the basic features of
the above-mentioned energy dependence, the most elementary
explicit toy model may be used and studied. Once we  contemplate
the Schr\"{o}dinger equation in one dimension (in units
$\hbar = 2$) with the harmonic oscillator interaction,
 \be
 -\,\frac{1}{m(E)}\ \frac{\rm d^2}{\rm dr^2}\, \Psi({r}) + {r}^2
\Psi({r})= E \Psi({r}) \label{hoSE}\
 \ee
the effects of the variability of the mass may be mimicked by an
arbitrary simulation  of its energy-dependence. For
the most elementary illustrative ansatz
 \be
 m(E) = A^2 \,(E-E_0)^2\,,
 \ee
a re-scaling of eq. (\ref{hoSE}) leads to the new
bound-state problem with the spectrum determined by the closed
formulae
 \be
 E=\left\{
 \ba
 E_n^{(+)}=E_0+\sqrt{E_0^2+\frac{8n+4}{A}}, \ \ \ \ n = 0, 1,
 \ldots,\\
 E_{\pm n}^{(-)}=E_0\pm \sqrt{E_0^2-\frac{8n+4}{A}}, \ \ \ \ n = 0, 1,
 \ldots,n_{max}
 \ea
 \right . \,.
 \ee
The latter two sets are finite and exist only for $A\,E_0^2\geq 4$.
The new levels emerge at each new $n_{max} =entier[(A\,E_0^2- 4)/8]$.
The message of this test is quite persuasive -- the structure of the
spectrum may be modified thoroughly even by a very innocent-looking
energy-dependent term.

\section*{Appendix B. An artificial
energy dependence originating from the model-space projections}

One of the most usual approaches to the realistic
Schr\"{o}dinger equation is variational. Typically, people start
from a suitable microscopic Hamiltonian.  Such a quantum model
is usually constructed on the basis of the correspondence
principle.  As a rule, it is energy-independent and complicated,
its handling proves time-consuming and its numerical
diagonalization appears distressingly slow.  These difficulties
force us to reduce the Hilbert space to a smaller subspace.
This means that we must replace the original microscopic
bound-state problem by its model-space version or reduction.  As
long as our attention is merely paid to the certain ``most
relevant" subsets of all the possible degrees of freedom, an
abstract algebraic reformulation of the above statement may be
based on a split of the identity operator $I$ into the projector
$P$ (on the manageable subspace) and its complement $Q=I-P$. The
exact and complete Schr\"{o}dinger equation $H \,|\Psi\rangle =
E\,|\Psi\rangle$ acquires the two-by-two partitioned form
 \ben
 P\,(H-E)\,P\,|\Psi\rangle +
 P\,(H-E)\,Q\,|\Psi\rangle =0,
\een
 \ben
 Q\,(H-E)\,P\,|\Psi\rangle +
 Q\,(H-E)\,Q\,|\Psi\rangle =0.
\een
The energy-dependence emerges when we eliminate the component $
Q\,|\Psi\rangle$ of the wave function from the second row and
insert it in the first equation.  The resulting explicit form of
the reduced problem,
 \ben
 P\,(H-E)\,P\,|\Psi\rangle +
 P\,H\,Q\,
\left [ \frac{Q}{Q\,(H-E)\,Q} \right ] \, Q\,H\,P\,|\Psi\rangle
  =0
\een
may be understood as a projected or ``effective" Schr\"{o}dinger
equation
 \be
H_{eff} \,|\Psi_{eff}\rangle = E\,|\Psi_{eff}\rangle\,.
\ee
The value of the energy remains unchanged while the wave
function is reduced to a subspace, $|\Psi_{eff}\rangle =
P\,|\Psi\rangle$. In general, the energy dependence appearing in
the effective Hamiltonian $H_{eff}(E)$ is very complicated.

%\newpage

\section*{Appendix C. Partial solvability re-interpreted as an
 energy-dependence }

For the sake of definiteness, let us pick up the two most elementary
and popular QES examples, viz., the Hautot's \cite{Hautot} shifted
harmonic oscillator defined at certain exceptional charges $F$ only,
 \be
 \left [
 -\frac{d^2}{dr^2} + \frac{\ell(\ell+1)}{r^2}
 + \frac{F}{r} + f\,r+ r^2
 \right ]\,\psi_{n,\ell}(r) = E_{n,\ell}\,\psi_{n,\ell}(r)
 \label{charged}
 \ee
and the sextic oscillator of Singh et al \cite{Singh} with certain
exceptional ``spring" constants $A$,
 \be
 \label{sextic}
 \left [
 -\frac{d^2}{dr^2} + \frac{\ell(\ell+1)}{r^2}
 + A\,r^2 + a\,r^4 + r^6
 \right ]\,\phi_{n,\ell}(r)
 = \varepsilon_{n,\ell}\,\phi_{n,\ell}(r)\,.
 \ee
Both these models may be made mathematically completely
equivalent via a suitable change of the variables \cite{classif}
but each of them plays a slightly different role in physics. We
may index their bound states in the same manner, viz., by their
angular momenta $\ell$ and by the number $n$ of nodes in the
wave function.

In both cases, the essence of the QES construction lies in the
requirement that the Taylor series for the bound-state wave functions
terminate and become proportional to a polynomial of degree $N$. In
the former, Coulombic case (\ref{charged}) this implies  that the
exceptional QES energy becomes fixed and remains unique and
charge-independent. Thus, the spectrum will be numbered by $N$ and
coincides, incidentally, with the equidistant energies of the pure
harmonic oscillator.  In contrast, at each $N$ there exist as many as
$N+1$ different QES-compatible values of the charge $F=F_{(N,j)}$,
with $j = 0,1,  \ldots, N$ (see \cite{Hautot} for details).  At each
energy, the admissible value of the charge becomes unique only after
a particular choice of the index $j=j_0$ which may vary with the
changes of the second index $N$. Thus, the QES charge may be
understood as an energy-dependent quantity, $ F=F_{QES}(E)$.

In the latter, sextic illustrative example (\ref{sextic}), the roles
of coupling and energy become interchanged \cite{classifb}. At any
angular momentum $\ell$, the main quantum number $N$ now counts the
eligible couplings $A=A_N$. The related $(N+1)-$plets of the
admissible energy values $\varepsilon_{n,j}$ must be generated by the
specific algorithm of ref. \cite{Singh}. Although the resulting
overall energy-dependence pattern remains very similar, its practical
aspects become less comfortable since up to the exceptional
large$-\ell$ limit \cite{Gerdt}, the correspondence between $E$ and
$N$ acquires an ackward numerical character at the larger $N$.

\end{document}